# Dragonflies Utilize Flapping Wings' Phasing and Spanwise Characteristics to Achieve Aerodynamic Performance

**Authors:** Csaba Hefler[1*], Huihe Qiu[1*], Wei Shyy[1*]

**Affiliations:**

[1] Department of Mechanical and Aerospace Engineering, The Hong Kong University of Science and Technology, Hong Kong SAR, China. (Address: Room 2568, Main Academic Building, Hong Kong University of Science and Technology, Clear Water Bay, Kowloon, Hong Kong)

*Correspondence to:  chefler@ust.hk  or  meqiu@ust.hk  or  weishyy@ust.hk (Address: Room 2568, Main Academic Building, Hong Kong University of Science and Technology, Clear Water Bay, Kowloon, Hong Kong, Tel.: +852 23587190)

**Abstract**: While dragonflies are highly agile flyers, some key aerodynamic mechanisms responsible for their flight performance remain inadequately understood. Based on forward flight conditions, we investigate dragonflies's spanwise aerodynamic behaviors associated with flapping wings' phasing relationship. Overall, the leading edge vortex (LEV) on the forewing forms without the influence of the hindwing. For hindwing, the wing root region prominently displays a trailing edge vortex (TEV). In the inner span region, the vortical flow structures around the hindwing is influenced by the forewing's LEV when both wings are in close proximity and move in opposite directions. In the mid-span region, downwash following the forewing suppresses LEV formation on the hindwing. Finally the outer span region of the hindwing develops its LEV by wake capture at the end of a stroke cycle. In the inner region, the timing of shedding on both fore- and hind-wings is synchronized, which is not the case elsewhere. These varied flow structures suggest that the fore- and hind-wings, along their spanwise directions, play different roles in force generation.

**Significance Statement.** A new insight into flapping wing's phasing and aerodynamic interactions in support of dragonflies' exceptional flight performance is reported. The first direct observation of a living dragonfly utilizing phasing and spanwise characteristics to control the flapping wing aerodynamic performance is investigated. It is found that a dragonfly can utilize its hindwing to capture its forewing's wake at the end of a stroke cycle so that the hindwing's LEV can be enhanced. While both biologists and engineering researchers have been joining force to unlock the secret of flapping wing aerodynamics and to design bio-inspired flying machines, this paper offers new understanding of high aspect ratio wing performance and how to create new design concepts.

\Body

**Introduction:** There are nearly a million species of flying insects. Of the non-insects, another 13,000 warm-blooded vertebrate species, including about 9,000 birds and 1,000 bats, have taken to the skies [1-2]. The insects, based on their sizes and flight speeds, operate in the low Reynolds number regime where both inertia and viscous effects are important, which, coupled with unsteady flexible flapping wing motion, results in a number of phenomena qualitatively different from the aerodynamic phenomena associated with fixed-wing flyers [1-4]. While the key unsteady aerodynamic mechanisms responsible for lift enhancement are well documented [5], many issues regarding wing-wing interactions, flapping kinematics, structural flexibility and wing-body synchronization are topics in need of substantial new insight [6-8]. Dragonflies, one of the fastest insects, exhibit fascinating flying skills, including gliding, powerful ascending, backwards flight and effective level flight. Their long and slim body with movable tail and two pairs of high aspect ratio flexible wings, in collaboration with highly evolved sensory capabilities results in a finely tuned system making them one of nature's most successful aerial hunters [9-11].

Interactions between the fore- and hind-wings play important roles in dragonflies' flight performance. During forward flight, dragonflies adopt wing phasing where the hindwing leads the forewing in flapping motion by about 90 degrees, which is much reduced if excess forces is needed when, e.g., they try to escape from captivity [19-25]. Downwash generated by the forewing is known to decrease the hindwing's effective angle of attack and possibly suppressing the leading edge vortex (LEV) formation [12-16] and lift generation [14]. It is shown that thrust is higher when the hindwing leads the flapping cycle [13, 16-18]. We also know that vortex interactions play important roles aerodynamically for tandem wing arrangements, as studied in, e.g., [16-18, 25-32]. For insects such as dragonflies and hawkmoths, operating at Reynolds number around ($10^3$–$10^4$), a leading edge vortex (LEV) is generated from the balance between the pressure gradient, the centrifugal force, and the Coriolis force. The LEV generates a lower pressure core which enhances lift. It is established that constructive vortex interaction can enhance the hindwing leading edge vortex generating additional aerodynamic force [17-18, 25-29]. A study with rigid wings modelled after dragonflies has found that constructive LEV interactions between forewing and hindwing can make the tandem configuration outperform wings operating in isolation [26]. Furthermore, flexible wings can vary the strokewise gap between the wings that affects vortex shedding and further influence force generation by modifying interaction dynamics [27, 28]. A vortex shed by the forewing can also reposition the hindwing's trailing edge vortex (TEV) to boost thrust by forming a jet between a counter rotating shed vortex pair [18] or to enhance hindwing lift by its low pressure core as it moves over the hindwing [16]. On the other hand, destructive vortex interactions can reduce the aerodynamics forces [26-28, 30-32]. In short, in-phase flapping can generate higher aerodynamic forces, while a higher aerodynamic efficiency is observed if the hindwing leads the forewing in flapping cycles. While the hindwing derives force from the wake behind a forewing, the higher aerodynamic efficiency is likely facilitated by a converging momentum flow with reduced swirl [34].

Considering dragonflies' high aspect ratio wings, it is unclear how the flapping phasing relationship between fore- and hind-wings influences the spanwise aerodynamics [6-8]. In this study, issues associated with such wing-wing interactions are investigated using time-resolved flow measurements with a particular focus on spanwise variations. Our goal is to better understand how dragonflies' hindwings utilize the fluid flows induced by the forewing from previous and present flapping cycles in time- and space-dependent manners to attain their exceptional flight performance. We have measured the flow field around the wings of a live dragonfly (Pantala Flavenscens), commonly known as wandering glider (figure 1.) based on time resolved particle image velocimetry (PIV).

In the experiment, a live dragonfly is glued to the tip of a glass plate that doesn't restrict the wing and tail motions. The body line angle is initially set to be 22 degrees inclined downward in order to support a favorable field of view for the PIV cameras. The flapping frequency of the dragonfly is measured to be 24 Hz. The flapping amplitudes for the fore- and hind-wings are around 65-70 degrees and 66-68 degrees, respectively. The stroke plane is the plane defined by the wing root and the pterostigma, a colored cell in the outer wing, at the stroke extremes, while the stroke plane angle is the angle between the stroke plane and the horizontal axis at the wing root fixed coordinate system. The stroke plane (taken the horizontal axis as reference) is inclined by around 83 degrees for both wings (figure 2). As illustrated in figure 2, the wing phasing kept around 80 degrees led by the hindwing. Assuming the deviation of the stroke plane angle is negligible, the distance between the fore- and hind-wing leading edges throughout a flapping cycle is defined by a geometrical wing sweeping perpendicular to the stroke plane (the forewing is forward swept, while the hindwing is backward swept) that doesn't change throughout the flapping cycle, and a varying strokewise distance due to the wing phasing. Resulting from an offset of the fore- and hind-wing flapping amplitudes there is an asymmetry of the angle between the leading edge of the wings projected to a common stroke plane in up- and down-strokes (figure S1). This results in a time dependent distance between the leading edges, as well as the forewing trailing edge and hindwing leading edge that is asymmetric trough the up- and down-strokes, account for an asymmetry of aerodynamic characteristics. Time resolved particle image velocimetry (TR-PIV) at a 1 kHz sampling rate was used to acquire particle image data along a vertical plane parallel to the dragonfly body. According to the momentum balance, the thrust needed for forward flight is accompanied by horizontally oriented flows. Our evaluation focuses on sequences when the tethered dragonfly kept generating a horizontal momentum for several flapping cycles that can be considered as forward flight mode. Additionally we have measured the flow field around the wings of a dragonfly in the outer span region in case of unrestricted take off flight that closely relates to the forward flight mode. Accordingly, a characteristic interaction of the outer spanwise region, qualitatively consistent with the interaction found for the forward flight mode, was recorded (video S2). The pterostigma trajectories and stroke amplitudes (figure 2) over time are assessed from the PIV frames based on the measurement on a plane located at 78 % span from the wingroot.

**Discussion:** Typically, the forewing develops an attached LEV during each stroke on its own, which rolls up and then is shed downstream, influencing the wake and the TEV. Depending on the distance of separation between the forewing's trailing edge and the hindwing's leading edge, and the flow structures associated with the wing motions and deformation during the stroke, multiple distinct patterns are observed along the spanwise direction.

In this paper, we focus on the following distinct regions, namely the root (the innermost 15% of the span), inner (approximately from 15 % to 45 % span), intermediate (approximately from 45% to 62% span) and outer (approximately from 62% span toward the wing tip) regions. Depending on the pairing conditions, including proximity between wings, relative motion, and surrounding flow structures, we observe various scenarios along the spanwise direction. Varied unsteady aerodynamic scenarios appear in different spanwise regions and phases during a flapping cycle. Compared to an in-phase tandem wing flapping, dragonflies' ingenious flapping wing operation during forward flight takes advantage of the proximity and movement between fore- and hind-wings, as well as the flow field created by the individual wings during present and previous stroke cycles to form a highly complex system for lift as well as thrust generation.

The main flow features related to the formation and shedding of LEV and development of TEV and subsequent vertical flows, associated with both fore- and hind-wings are discussed below.

**Root region (video S3):** Due to the combined wall effect resulting from the body and the small flapping amplitudes, insignificant aerodynamic performance is associated with the wing root region. The leading edge of the forewing guides and distributes the incoming flow while forming a weak LEV. The trailing edge of the hindwing creates a noticeable TEV while it translates with noticeable pitch angle variations. The wings show a similar behavior as wings flapping in-phase; there is no noticeable vortex interaction between the wings.

**Inner span region (Figure 8, video S4):** As the hindwing leads the forewing in a flapping cycle, at certain instants the two wings move in opposite direction with the leading edge of the hindwing in close proximity of the trailing edge of the forewing. The opposite wing movement and the forewing LEV together facilitate a LEV formation on the hindwing. As the stroke cycle progresses, the timing of shedding on both fore- and hind-wings is synchronized. As a result of this interaction the hindwing LEV extends and covers a larger area of the hindwing surface. This intensification of the hindwing LEV can produce additional aerodynamic force [17, 25-26]. In the mid-span and outer span regions, the gap between fore- and hind-wings is larger which diminishes the synergic interactions between the fore- and hind-wing's LEVs.

**Mid-span region (Figure 8, video S5):** The forewing forms an LEV during the course of the stroke cycle, which is shed when the wing rotates during stroke reversal. In the process, the tip of the forewing is curved toward the direction of wing movement, creating a concave camber enveloping the vortex. The wing deformation along with its flapping motion influences the vortex formation and subsequent shedding. The resulting downwash influences the effective angle of attack of the hindwing and suppresses the formation of a coherent LEV on the hindwing. Hence, the mid-span section of the hindwing largely redirects the flow and the direction of the associated aerodynamic force.

**Outer span region (Figure 8, video S6):** In the outer span region the forewing operates similarly as in the mid-span. It generates a single LEV in each stroke which is shed after the rotational motion during the stroke reversal phase. This vortex experiences a circulation boost from the trailing section of the translating forewing, and preserves its integrity well. The hindwing operates under the influence of a downwash behind the forewing. The LEV on the hindwing is formed only toward the later stage of a stroke cycle. In particular, the separation distance between the fore- and hind-wing's outer regions and their flapping phasing relationship together allow the hindwing to interact with the forewing vortex shed during the previous stroke. The hindwing captures this shed vortex by its suction surface to form an enhanced LEV during the last quarter of the stroke cycle. This interaction provides additional aerodynamic force for the hindwing in a later phase of its stroke cycle.

In summary, the distinct regions for different main flow features along the hindwing's spanwise direction are shown in figure 4. The wake captures only appear at the end of strokes in the hindwing's outer-span region. The hindwing amplitudes for up- and down-stroke vortex interactions are asymmetric due to the strokewise offset between the flapping amplitude of the fore- and hind-wings. The LEV formation of the forewing takes place as it rotates and translates, while that of the hindwing is substantially influenced by the forewing. In the inner span region, the hindwing's LEV is formed and shed in response to the forewing's close proximity and movement. Shedding synchronization of both LEVs is observed. In the mid-span region, the hindwing's LEV is less coherent and the flow and force redirection is an important feature of the hindwing there. For the outer span region, wake capture is a major force enhancement mechanism for the hindwing. Overall, the forewing is surrounded by coherent vortical flows during much of a flapping cycle, but the hindwing exhibits such flow patterns mainly toward the end/beginning of a stroke. These varied flow structures suggest that the fore- and hind-wings, along their spanwise directions, play different roles in lift and thrust generation.

As previously stated, the distance between wings during up- and down-strokes is asymmetric because the stroke angles of the fore- and hind-wings are different. While the main flow features described above are valid for both strokes, the detailed characteristics change. Based on these observations, the LEV's development and shedding from the forewing of a forward flying dragonfly is different from that of a single flapping wing or wings with in-phase motion. Without pressure variations created by the hindwing the LEV is reported to break down in the middle of the downstroke [35]. No such vortex breakdown is observed in our study.

Figure 5 offers three-dimensional perspectives of wing-wing and wing-flow interactions, depicting both attached LEVs and TEVs as well as those shed recently. The first one corresponds to the instant when the hindwing is at mid-course, with the LEV being present mainly in the inner span region. The vortical flow in the hindwing's mid-span region is less than coherent throughout a flapping cycle, and that in the outer region hasn't been formed yet. In the second plot, corresponding to the instant slightly before the hindwing reaches the end of a stroke, wake capture takes place in the hindwing's outer span region.

Finally, to complement the observed flow features, we note that the wings curve substantially while they rotate and turn, indicating the influence of the inertia and centripetal forces. As illustrated in figure 5, the aerodynamic forces also modify the shape in the course of flapping movement. While the wings in general tilt toward the direction of movement, their cross-sectional shapes vary along the spanwise locations due to the combination of the inertia, the pressure distributions and the corrugated wing structures [36-37]. We note that the structural flexibility of corrugated wings clearly adds further complexity to dragonflies' flight performance. The flexible, flapping wing dynamics is a subject being actively studied [2, 38, 39], and expected to be further investigated.

**Acknowledgement:** This research is supported by Hong Kong Ph.D. Fellowship Scheme from the Research Grants Council (RGC), the Government of the Hong Kong Special Administrative Region (HKSAR) and Hong Kong University of Science & Technology.

**Figure Legends.**
> **Figure 1** Image of a wandering glider (Pantala Flavenscens)
> (Source: Hong Kong Wetland Park of the Agriculture, Fisheries and Conservation Department)
> **Figure 2:** Stroke amplitude for two flapping cycles and the pterostigma trajectory for one cycle
> **Figure 3:** Main flow features in different spanwise positions, that affecting the hindwing at it's doiwnstroke (* denotes vortexes that formed in the previous stroke cycle)
> **Figure 4:** Main flow features along the hindwing's spanwise direction
> **Figure 5:** 3D schematic of the flow features at selected time instants. (* marks the vortical structures formed in the previous cycle)

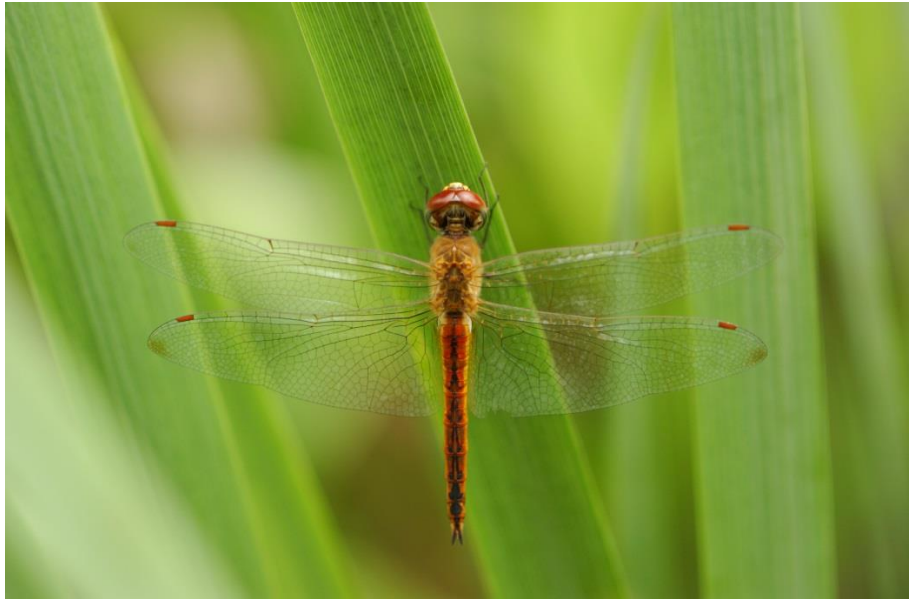

**Figure 6** Image of a wandering glider (Pantala Flavenscens)
(Source: Hong Kong Wetland Park of the Agriculture, Fisheries and Conservation Department)

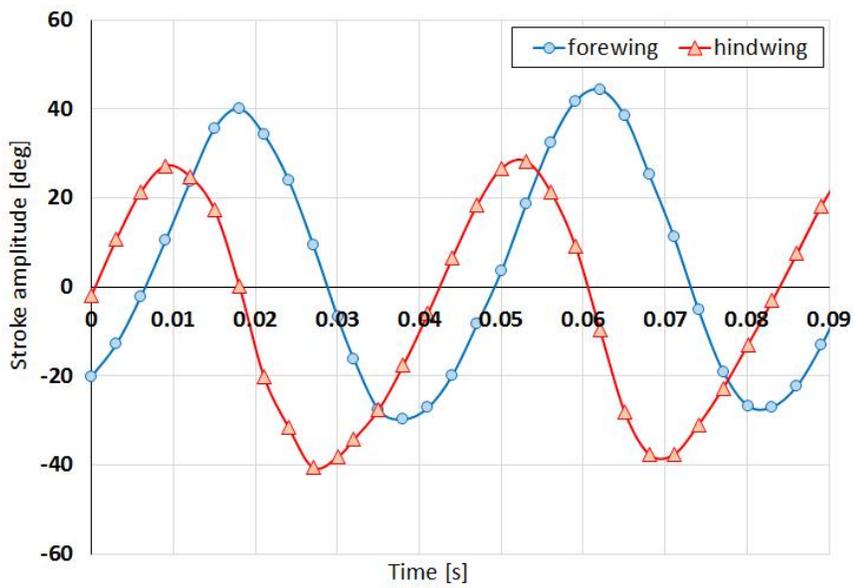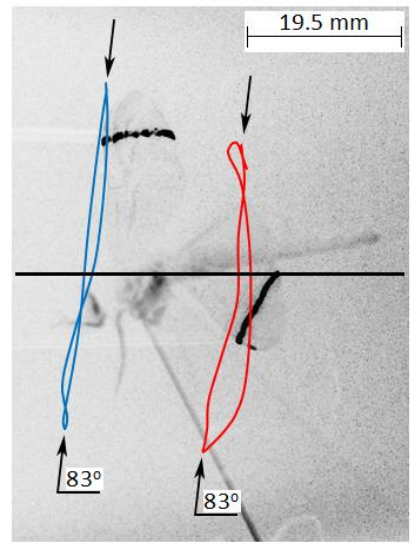

**Figure 7:** Stroke amplitude for two flapping cycles and the pterostigma trajectory for one cycle

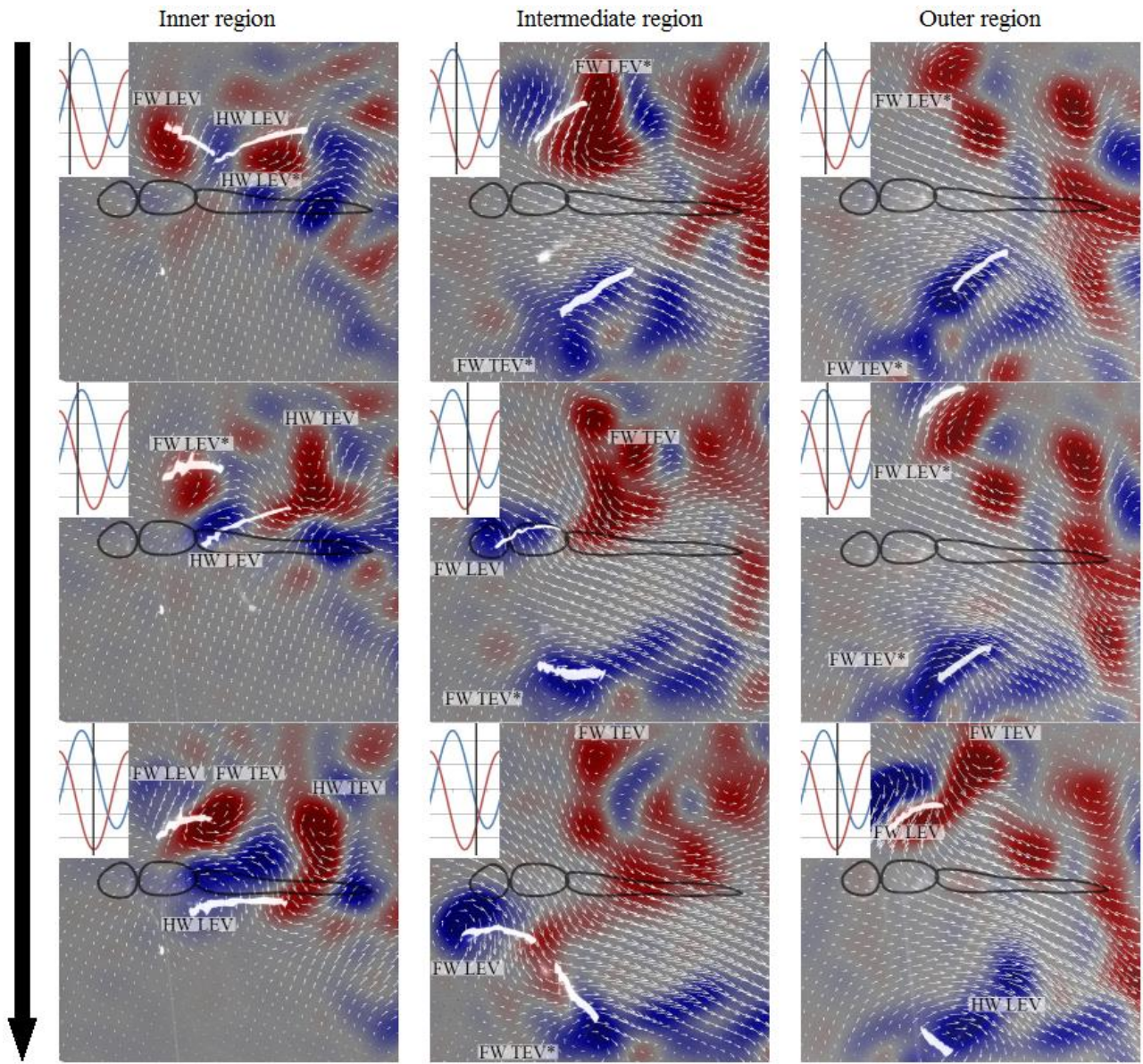

**Figure 8:** Main flow features in different spanwise positions, that affecting the hindwing at it's doiwnstroke (* denotes vortexes that formed in the previous stroke cycle)

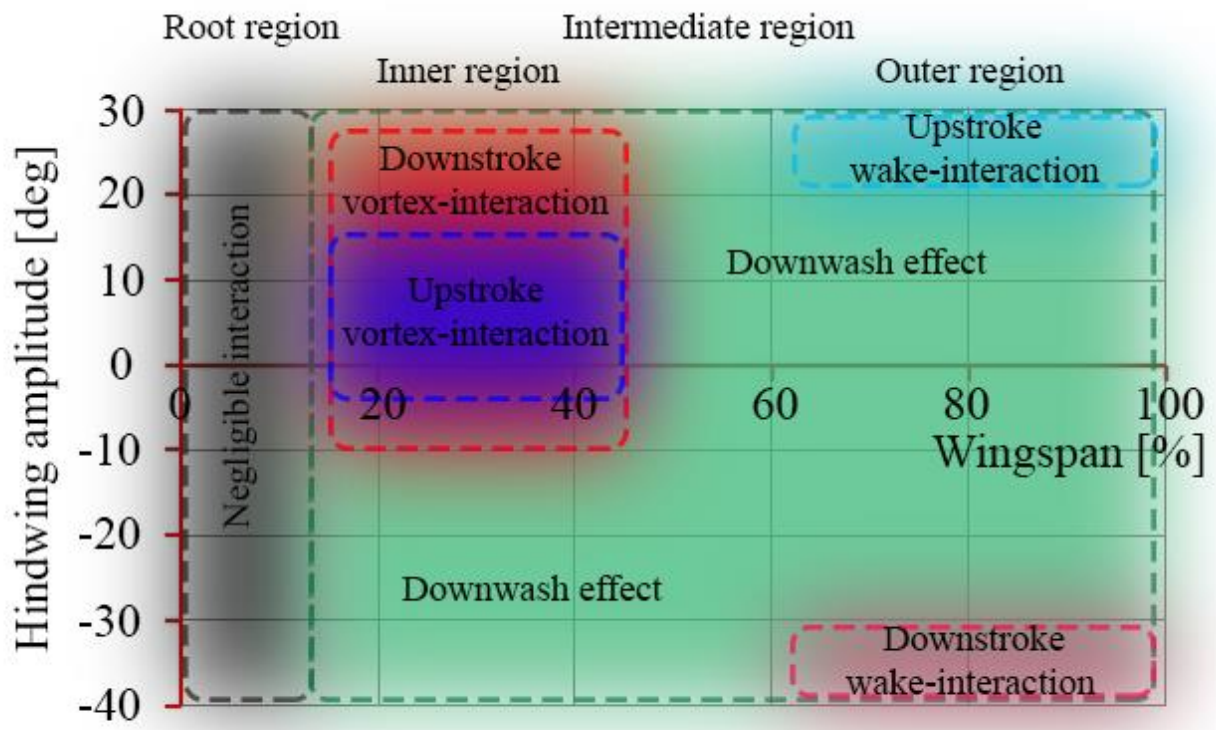

**Figure 9:** Main flow features along the hindwing's spanwise direction

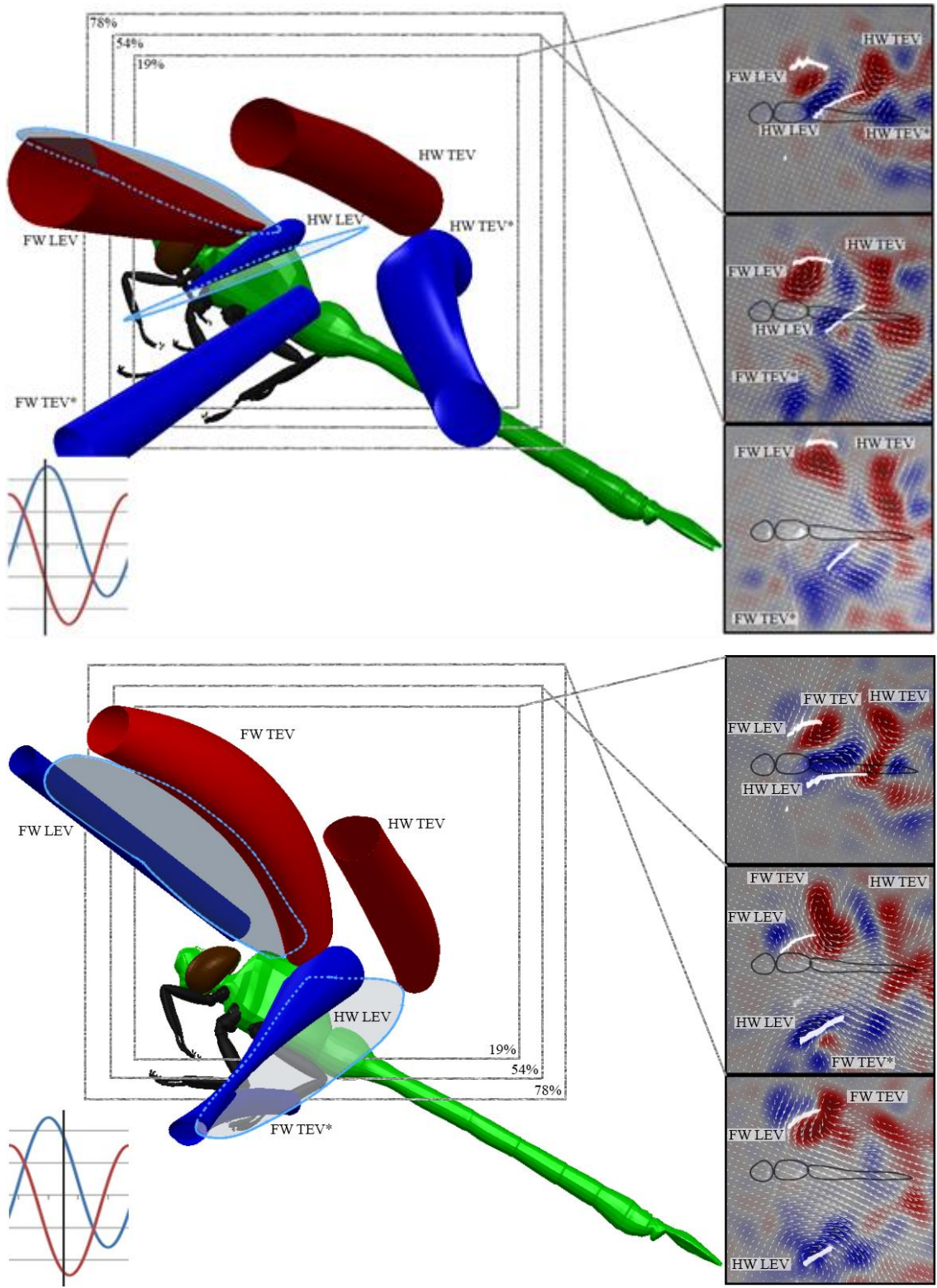

**Figure 10:** 3D schematic of the flow features at selected time instants. (* marks the vortical structures formed in the previous cycle)

**Supplementary Materials:**

Materials and Methods

Figures S1 (Not released)

Video S2 (Not released)

Video S3 (Not released)

Video S4 (Not released)

Video S5 (Not released)

Video S6 (Not released)

Author contributions

**Supplementary materials:**

**Materials and methods:**

**Catching dragonflies:** Pantala Flavenscens, a very common species (not protected) in Hong Kong, was caught near HKUST campus. It is a medium sized dragonfly with a wingspan of 80-90 mm, and body length of about 50 mm. The dragonflies were used within a few hours after capture; allowing sufficient time for the specimen to adapt to the lower temperature in the laboratory.

**Tethered dragonflies:** The dragonfly was glued to a transparent glass plate with transparent epoxy glue at its thorax. While being tethered, the dragonfly's wings and tail motions are not restricted. In this study, the forward flight mode is defined as when the wings generate horizontal momentum fluxes over multiple flapping cycles. The angle between the body line of the dragonfly and the horizontal axis was 22 degrees. The 1.1 mm thick glass with narrowing sides towards the dragonfly abdomen was rigid enough to eliminate vibrations and its transparency ensured minimal glare. The glass plate was held by a precision stage allowing the specimen to be moved precisely in the spanwise direction.

**Locations of measurements:** The measurement was done at several spanwise positions along the spanwise axis (7%, 19%, 31%, 42%, 54%, 66%, 78% wingspan considering the laser light sheet at the hindwing midstroke position). As the wings flapped, the span wise position changed accordingly, but for laser light sheet operated at a high sampling frequency of 1000 Hz this change was negligible between the consecutive frames. Accordingly, the wing flapping features and the induced flow structures did not change when the measurement position changes. The dragonfly was stimulated with a thin carbon fiber rod. To minimize disturbances to the flow caused by the rod, the specimen was first allowed to grab the rod akin to perching on a tree branch. Then the rod was pulled away from its grasp, causing it to instinctively start to flap.

**Velocity capture system:** To measure the flow field around the flapping wings, a time resolved stereo particle image velocimetry (PIV, LaVision) was used. Flow field areas of approximately 130 x 70 mm were recorded using two high-speed CMOS-sensor cameras (VC-Phantom M310). The cameras were equipped with 85 mm lenses (MF Nikkor 24-85 mm, f/2.8) mounted on Scheimpflug adapters. The closed air reservoir was seeded with a mist of olive oil from a compressed air aerosol generator (LaVision UK Ltd). The seeding particles were of submicron diameter. Before each measurement time was given for the reservoir to calm from the induced flow by the seeding pump. The oil mist was illuminated by a 20 mJ laser (VL-Nd: YLF, 527 nm, Photonics Industries, US) producing double pulses with a repetition rate of 1000 Hz and a pulse separation interval of 200 µs. The laser beam was delivered by a compact light guiding arm (LaVision GmbH), the light sheet was approximately 2 mm thick parallel to the dragonfly's body. The laser and the PIV cameras were controlled by DAVIS v. 8.2.1 software package and were triggered by a common high-speed controller. The PIV cameras were calibrated with the built-in calibration routine in DAVIS v. 8.2.1 using a 310 x 310 mm dual plane calibration plate (LaVision, type 309-15). The calibration than was refined using a built-in auto calibration algorithm by two steps. The recorded PIV data were processed using DAVIS v. 8.2.1. Vector fields were computed from the filtered images using stereo cross correlation mode starting with interrogation windows of 96 x 96 pixels, reducing stepwise to 32 x 32 pixels for the final pass. Vectors were considered erroneous and deleted by the applied median filtering if the magnitude was equal to or more than three times the neighborhood root mean square (r.m.s.) and reinserted or changed to second or third choice vectors if the magnitude was less than twice the remaining neighborhood r.m.s. (closest neighbor vectors). Empty spaces were filled by interpolation (an average of all non-zero neighbor vectors) and the final vector fields were subject to a 3 x 3 smoothing.

**Visualization system:** Additional high speed video recording was executed on another specimen of the same dragonfly species to evaluate the differences in wing spacing in the up- and down-strokes. The dragonfly was fixed similarly to a glass plate, with a body line alignment that resulted vertical stroke planes. The camera was positioned perpendicular to the stroke plane of the wings. The dragonfly was stimulated using a carbon fiber rod similarly to the above described flow measurement. To record the wing motion a Redlake MotionXtra HG-100K high speed camera with MF Nikkor 50mm, f/1.4 was used with a frame rate of 2000 frames per seconds.

**Data processing and analysis:** All data processing and analysis was performed on a computer. A sufficient number (For each experiment, at least 3 flight measurements at each section for an individual specimen; in each flight measurement, at least 30 flapping cycles were recorded by PIV and high speed camera.) of dragonflies and flights were collected for each condition such that all results could be reproduced robustly. No statistical method was used to predetermine sample size.